\documentclass[10 pt]{article}
\usepackage{epsfig}
\begin{document}

\title{\bf Quantum Destructive Interference}

\author{A.Y. Shiekh\footnote{\rm shiekh@dinecollege.edu} \\
             {\it Din\'{e} College, Tsaile, Arizona, U.S.A.}}

\date{}

\maketitle

\abstract{An apparent paradox for unitarity non-conservation is investigated for the case of destructive quantum interference.}

\baselineskip .5 cm

\section{Introduction}
Destructive quantum interference might at first seem to potentially threaten unitarity since it seems to imply a reduction of the wave-function. Since this mechanism has been exploited for a variety of applications \cite{Shiekh1, Shiekh2, Shiekh3}, this motivates a deeper investigation into the seeming paradox of unitary conservation.

\section{Quantum Interference}
An extreme example of destructive quantum interference is where two arms are brought into overlap, having first arranged for them to be in anti-phase

\begin{center}
\includegraphics[scale=.5]{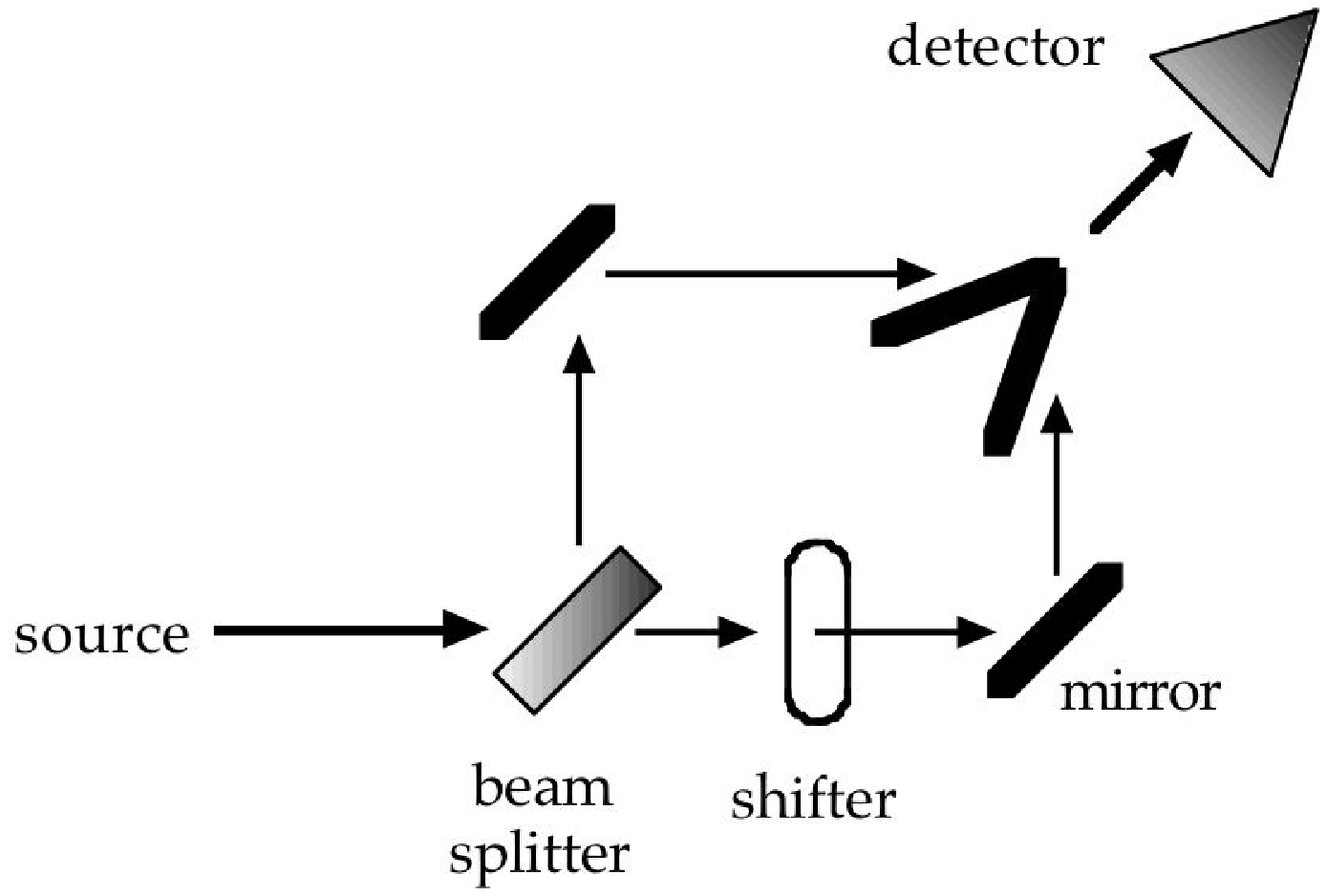}
\end{center}
\begin{center}
Destructive Interference
\end{center}

In this case the destructive interference can be near complete in practice, and one might naturally wonder if this is in conflict with unitarity conservation.

\subsection{Multi-valued wave-`functions'}
As is well known, expressions such as the square root are not functions as there is, in general, more than one result. In response to this dilemma one can either force the issue by defining the principle square root to be only one of the two valid outputs, or one can restore the single valuedness by defining it on a two sheeted Riemann surface.

\begin{center}
\includegraphics[scale=.5]{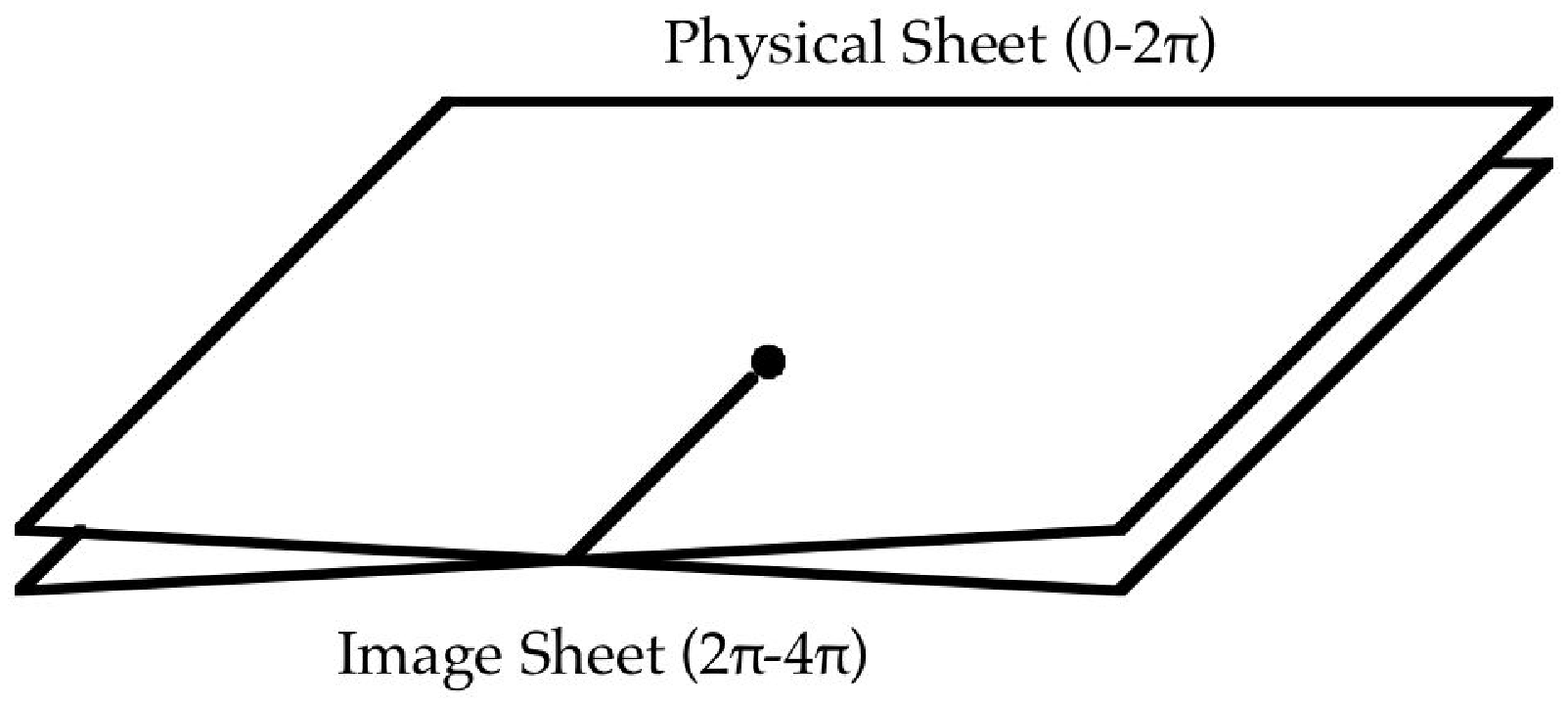}
\end{center}
\begin{center}
Riemann double sheet
\end{center}

The interference situation being considered here is similar to the Aharonov-Bohm effect, and also lives on a Riemann surface due to the multi-valued nature of the situation; in this case the same two sheeted surface as for the square root.

Now the interfering arms on the physical (top) sheet combine destructively as
$$
(\left| \psi \right> - \left| \psi \right>)/2
$$
while on the lower (unphysical) sheet they combine constructively as
$$
(\left| \psi \right> + \left| \psi \right>)/2
$$
so preserving unitarity over the entire mathematical space; the normalization of $1/2$ is composed of $1/\sqrt{2}$ for each path, and $1/\sqrt{2}$ for each sheet.

More generally, for a phase shift of $2 \pi \frac{n}{m}$ ($n$ and $m$ integers; $m$ strictly positive), the sum of probabilities over the now $m$ sheets becomes
$$
\sum_{l=1}^{m} \left| \frac{1}{\sqrt{2m}} \left(  1 + e^{2\pi i l \frac{n}{m}} \right) \right|^2
$$
which when multiplied out yields
$$
\frac{1}{2 m}\sum_{l=1}^{m}  \left( 2 + e^{-2\pi i l \frac{n}{m}} + e^{2\pi i l \frac{n}{m}} \right)
$$
and is then seen to be identically equal to one since each exponent is the sum of vectors spread evenly around zero. This demonstrates that unitarity is indeed mathematically preserved over the whole Riemann surface.

\section{Conclusion}
The seeming paradox of lost unitarity is resolved by the realization that mathematically, over the whole Riemann surface, unitarity is actually preserved; while destructive interference can take place on the single physical sheet.

However, since the particle cannot actually be lost, the wave-function needs to be renormalized on the physical sheet, and this constitutes the mechanism that is being exploited for quantum computation and communication \cite{Shiekh1, Shiekh2, Shiekh3}.

The use of Riemann surfaces to carry physical problems can be of great utility \cite{Shiekh4}.

\end{document}